\documentclass[a4paper]{jpconf}
\usepackage{graphicx}
\usepackage{iopams} 
\usepackage{graphics}
\usepackage{sidecap}
\usepackage{rotating}
\usepackage{tikz,ifthen,calc}
\usetikzlibrary{decorations.markings}
\usetikzlibrary{plotmarks}
\usetikzlibrary{arrows,matrix}
\begin{document}
\title{Competition of different coupling schemes in atomic nuclei}

\author{C. Qi, R. J. Liotta, R. Wyss}

\address{KTH, Alba Nova University Center, SE-10691 Stockholm, Sweden}

\ead{chongq@kth.se}

\begin{abstract}
Shell model calculations reveal that the ground and low-lying yrast states of the $N=Z$ nuclei $^{92}_{46}$Pd and $^{96}$Cd are mainly built upon isoscalar spin-aligned neutron-proton pairs each carrying the maximum angular momentum $J=9$ allowed by the shell $0g_{9/2}$ which is dominant in this nuclear region.  This mode of excitation is unique
in nuclei and indicates that the spin-aligned pair has to be considered as an essential
building block in nuclear structure calculations. In this contribution we will discuss this neutron-proton pair coupling scheme in detail. In particular, we will explore the competition between the normal monopole pair coupling and the spin-aligned coupling schemes. Such a coupling may be useful in elucidating the structure properties of $N=Z$ and neighboring nuclei.
\end{abstract}

The low-lying structures of many nuclei can be understood as the outcome of the competition between the pairing (or seniority) coupling scheme and the aligned coupling of individual  particles  in  a  non-spherical  average  potential. The seniority coupling dominates the low-lying states of semi-magic nuclei, 
where the driving force behind is the strong pairing interaction between like particles.  Meanwhile, many open-shell nuclei have quadrupole moments that are  much larger than could be attributed to a single particle, which implies the sharing of angular momentum between many particles. A remarkable feature of nuclear structure physics is that essential ingredients of the single-particle model could be retained by assuming that the nucleons move in an average nuclear field deviates from spherical symmetry, which removes the spurious degrees of freedom corresponding to the collective spectrum \cite{Bohr76}. The transition between there two couplings in open-shell nuclei is expected to be rather fast \cite{Bohr80,Rowe98,Rosen07}.

Near the $N = Z$ line, nuclear collectivity
may be further enhanced by the interactions arising between neutrons and
protons occupying orbitals
with the same quantum numbers, which can form $np$ pairs with angular momenta $J=0$ to $2j$ and isospin quantum numbers $T=0$ (isoscalar) and $T=1$ (isovector). The isovector $np$ channel manifests itself in a fashion 
similar to like-nucleon correlations. There has been longstanding and intensive interest in exploring the signature 
of isoscalar $np$ pairing (in particular $np$ pairs with $J=1$ and $S=0$) in the structure of self-conjugate nuclei \cite{Goo79,Sat97,She00,duk06,Ler07,San09}. Possible fingerprints may be inferred from the delayed alignments in the 
ground-state rotational bands of deformed $N=Z$ nuclei, but it is still controversial. In this contribution we would like to discuss the so-called spin-aligned neutron-proton pair coupling from a nuclear shell-model perspective. In some spherical $N=Z$ nuclei it may overcome the normal pairing as the dominant coupling scheme of the ground state structure. We will show that the aligned $np$ pairs can generate striking regular evolution patterns in the energy spectra and transition probabilities along the yrast states of $N=Z$ nuclei. These may be deemed as a new kind of collective mode with isoscalar character that is unique in the atomic nucleus.
In a recent work the low-lying yrast 
states in $^{92}_{46}$Pd were observed and it was inferred that for the first time a transition 
from the isovector pair coupling mode to such spin-aligned $np$ coupling scheme may  have occurred~\cite{ced10,qi11}. For related earlier works, see Refs. \cite{dan67,dan66,bha71,wong71,chen92}.

\begin{figure}
  \begin{center}
\begin{tikzpicture}[scale=0.60, y=1.0cm,x=1.0cm,every text node part/.style={font=\small} ]


\def\gsposy{1.1}
\def\firstbandx{0.6}
\def\levelwidth{1.6}
\def\bandwidth{3.0}
\def\nuclidespace{0.60}
\def\yscale{0.0020}
\def\nucposy{19.5}

\def\yscale{0.0030}


\def\x{\firstbandx}
\def\y{\gsposy+\yscale * \energy}
\foreach \energy / \sp in {0/$0^+$,616/$2^+$,1416/$4^+$,2380/$6^+$, 3480/$8^+$}{
\draw[thick] (\x,\y) -- (\x+\levelwidth,\y);
\node[above] at (\x+\levelwidth,\y) {\small  \energy};
\node[above] at (\x,\y) {\sp};
}
\node[below] at (\x+0.5 * \levelwidth,\gsposy) {\large $^{88}$Ru};
\node[below] at (\x+0.5 * \levelwidth,\gsposy-1.0) {\large exp};
\def\ymax{\yscale * 3500}


\def\x{\firstbandx+1*\bandwidth}
\def\y{\gsposy+\yscale * \energy}
\foreach \energy / \sp in {0/$0^+$,567/$2^+$,1281/$4^+$,2031/$6^+$, 2804/$8^+$, 3648/$10^+$}{
\draw[thick] (\x,\y) -- (\x+\levelwidth,\y);
\node[above] at (\x+\levelwidth,\y) {\small \energy};
\node[above] at (\x,\y) {\sp};
\ifthenelse{\energy=878}{\node[below] at (\x+0.5*\levelwidth,\y) {\large \emph{15}};}{}
\ifthenelse{\energy=1708}{\node[below] at (\x+0.5*\levelwidth,\y) {\large \emph{20}};}{}

}
\node[below] at (\x+0.5 * \levelwidth,\gsposy) {\large $^{88}$Ru};
\node[below] at (\x+0.5 * \levelwidth,\gsposy-0.8) {\large fpg};

\def\x{\firstbandx+2*\bandwidth}
\def\y{\gsposy+\yscale * \energy}
\foreach \energy / \sp in {0/$0^+$,925/$2^+$,1792/$4^+$,2599/$6^+$,3212/$8^+$,4174/$10^+$}{
\draw[thick] (\x,\y) -- (\x+\levelwidth,\y);
\node[above] at (\x+\levelwidth,\y) {\small  \energy};
\node[above] at (\x,\y) {\sp};
}
\node[below] at (\x+0.5 * \levelwidth,\gsposy) {\large $^{88}$Ru};
\node[below] at (\x+0.5 * \levelwidth,\gsposy-1.0) {\large pg};

\def\x{\firstbandx+3*\bandwidth}
\def\y{\gsposy+\yscale * \energy}
\draw[thick, dashed, color=black!50] (\x-0.3,0) -- (\x-0.3,\gsposy+\yscale * 4100);


\def\x{\firstbandx+3*\bandwidth+\nuclidespace}
\def\y{\gsposy+\yscale * \energy}
\foreach \energy / \sp in {0/$0^+$,901/$2^+$,1964/$4^+$,2957/$6^+$,3404/$8^+$}{
\draw[thick] (\x,\y) -- (\x+\levelwidth,\y);

\ifthenelse{\energy = 2574}{}{\node[above] at (\x+\levelwidth,\y) {\small \energy};}
\ifthenelse{\energy = 2574}{\node[above] at (\x,\y-\yscale*120) {\sp};}{\node[above] at (\x,\y) {\sp};}
}
\node[below] at (\x+0.5 * \levelwidth,\gsposy) {\large $^{96}$Cd};
\node[below] at (\x+0.5 * \levelwidth,\gsposy-0.8) {\large fpg};


\def\x{\firstbandx+4*\bandwidth+\nuclidespace}
\def\y{\gsposy+\yscale * \energy}
\foreach \energy / \sp in {0/$0^+$,920/$2^+$,1971/$4^+$,2967/$6^+$,3349/$8^+$}{
\draw[thick] (\x,\y) -- (\x+\levelwidth,\y);
\ifthenelse{\energy = 2519}{}{\node[above] at (\x+\levelwidth,\y) {\small  \energy};}
\ifthenelse{\energy = 2519}{\node[above] at (\x,\y-\yscale*140) {\sp};}{\node[above] at (\x,\y) {\sp};}
}
\node[below] at (\x+0.5 * \levelwidth,\gsposy) {\large $^{96}$Cd};
\node[below] at (\x+0.5 * \levelwidth,\gsposy-1.0) {\large pg};

\def\x{\firstbandx+5*\bandwidth+\nuclidespace}
\def\y{\gsposy+\yscale * \energy}
\draw[thick, dashed, color=black!50] (\x-0.3,0) -- (\x-0.3,\gsposy+\yscale * 4100);

\def\x{\firstbandx+5*\bandwidth+2*\nuclidespace}
\def\y{\gsposy+\yscale * \energy}
\foreach \energy / \sp in {0/$0^+$,1021/$2^+$,1853/$4^+$,2273/$6^+$,2306/$8^+$}{
\draw[thick] (\x,\y) -- (\x+\levelwidth,\y);
\ifthenelse{\energy = 2273}{\node[above] at (\x+\levelwidth,\y-\yscale*220) {\small  \energy};}{\node[above] at (\x+\levelwidth,\y) {\small  \energy};}
\ifthenelse{\energy = 2273}{\node[above] at (\x,\y-\yscale*220) {\sp};}{\node[above] at (\x,\y) {\sp};}
}
\node[below] at (\x+0.5 * \levelwidth,\gsposy) {\large $^{98}$Cd};
\node[below] at (\x+0.5 * \levelwidth,\gsposy-0.8) {\large fpg};

\def\x{\firstbandx+6*\bandwidth+2*\nuclidespace}
\def\y{\gsposy+\yscale * \energy}
\foreach \energy / \sp in {0/$0^+$,1484/$2^+$,2280/$4^+$,2533/$6^+$,2674/$8^+$}{
\draw[thick] (\x,\y) -- (\x+\levelwidth,\y);

\ifthenelse{\energy = 2474}{}{\node[above] at (\x+\levelwidth,\y) {\small \energy};}
\ifthenelse{\energy = 2474}{\node[above] at (\x,\y-\yscale*140) {\sp};}{\node[above] at (\x,\y) {\sp};}
}
\node[below] at (\x+0.5 * \levelwidth,\gsposy) {\large $^{98}$Cd};
\node[below] at (\x+0.5 * \levelwidth,\gsposy-1.0) {\large pg};

\def\x{\firstbandx+7*\bandwidth+2*\nuclidespace}
\def\y{\gsposy+\yscale * \energy}
\foreach \energy / \sp in {0/$0^+$,1395/$2^+$,2082/$4^+$,2280/$6^+$, 2475/$8^+$}{
\draw[thick] (\x,\y) -- (\x+\levelwidth,\y);
\node[above] at (\x+\levelwidth,\y) {\small  \energy};
\node[above] at (\x,\y) {\sp};
}
\node[below] at (\x+0.5 * \levelwidth,\gsposy) {\large $^{98}$Cd};
\node[below] at (\x+0.5 * \levelwidth,\gsposy-1.0) {\large exp};

\end{tikzpicture}
\caption{Experimental \cite{Mar01,Nudat} and calculated yrast level schemes of $N=Z$ nuclei $^{88}$Ru and $^{96}$Cd and $^{98}$Cd with two proton holes inside the core $^{100}$Sn.}
\label{ru88}
\end{center}
\end{figure}
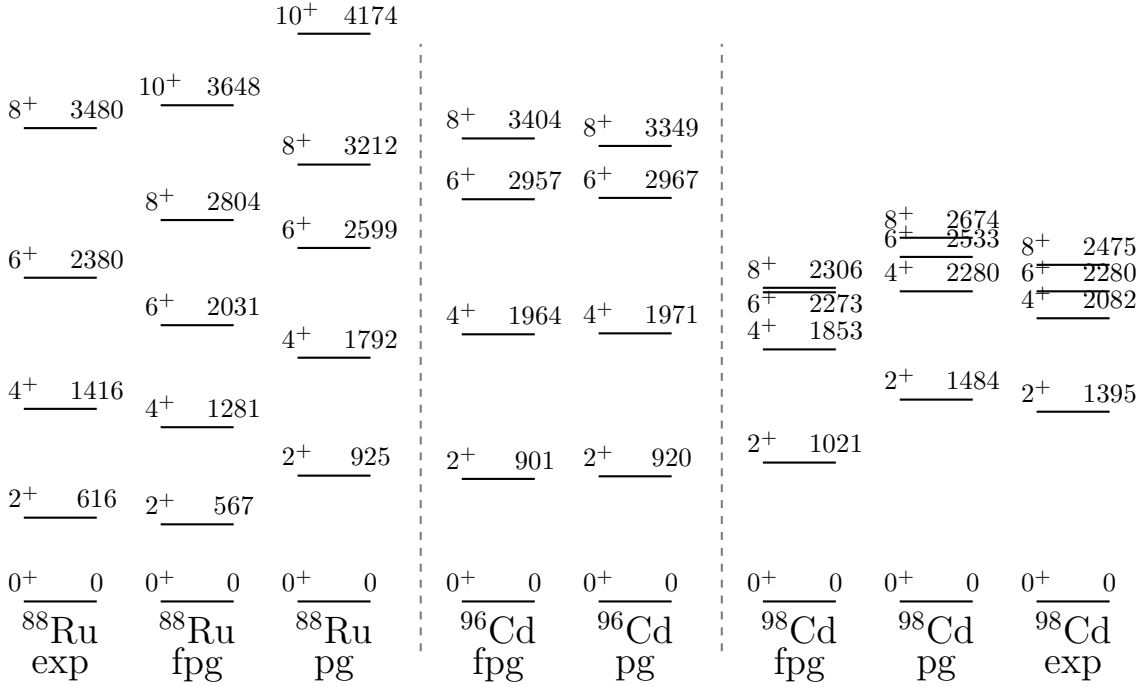

I would like to remind you that, as one would expect, near the closed shell nucleus $^{100}$Sn, i.e., in
$^{96}$Pd ($^{98}$Cd) with four (two) proton holes outside of $^{100}$Sn, 
the positions of the energy levels correspond
to a $(g_{9/2})_\lambda ^2$ pairing
spectrum, as can be seen from Fig. 1 of Ref. \cite{qi11} and Fig. \ref{ru88}.  When the number of 
neutron holes 
increases, the 
levels tend to be equally
separated, which is a characteristic of vibrational spectra. This are
specially the cases for $^{92}$Pd and $^{96}$Cd which show equally-spaced level schemes up to $I=12$ and $I=6$, respectively. 
Already the systematics of experimental data suggests gradual decrements in both
the quadrupole deformation $\beta_2$ and $E(4^+_1)/E(2^+_1)$ ratio in $N=Z$ nuclei when approaching the $^{100}$Sn shell closure \cite{Gar08,Mar01}. Many open shell $N=Z$ nuclei, like $^{88}$Ru shown in Fig. \ref{ru88}, exhibit rotational-like ground state bands. The nuclei $^{76}$Sr and $^{80}$Zr are the most deformed $N=Z$ ones in this region.

\begin{center}
\begin{table}[h]
\caption{The $0g_{9/2}$ interaction matrix elements for different model spaces \cite{hon09,joh01,Qi10}.\label{table}}
\centering
\begin{tabular}{@{}*{11}{c}}
\br
Space&$J=0$&$J=2$&$J=4$&$J=6$&$J=8$&$J=1$&$J=3$&$J=5$&$J=7$&$J=9$\\
\mr
$0g_{9/2}$&0&1.458&2.283&2.549&2.688&1.220&1.592&1.882&1.930&0.626\\
$pg$&-2.089&-0.919&-0.094&0.172&0.311&-1.038&-0.785&-0.495&-0.447&-1.751\\
$fpg$&-1.691&-0.959&-0.087&0.152&0.269&-1.138&-0.599&-0.383&-0.561&-2.207\\
\br
\end{tabular}
\end{table}
\end{center}

To understand the 
yrast structures of $^{92}$Pd, $^{96}$Cd and neighboring nuclei, we perform nuclear shell model calculations
within the $1p_{3/2}0f_{5/2}1p_{1/2}0g_{9/2}$ ($fpg$), $0g_{9/2}1p_{1/2}$
($pg$) and $0g_{9/2}$ model spaces using 
the Hamiltonian given in Ref. \cite{hon09} and a variety of other interactions which are quoted therein. In particular, we perform
calculations using as single-particle states the orbits 
$pg$ and $0g_{9/2}$, with the interactions of Refs. \cite{joh01,sch76,Qi10}, in order to explore the importance of configuration
mixing in determining the structure of the spectrum.

In Table \ref{table} we compare the strengths of the interaction matrix elements corresponding to different shell model spaces \cite{hon09,joh01,Qi10}. In the case of $0g_{9/2}$ space, only the relative strengths are given for simplicity. The absolute strength of the monopole centroid has no influence on the coupling of the wave functions and excitation energies. Also it should be mentioned that the relative strength of the $T=0$ and $T=1$ monopole interactions does not play any role on the structure of nuclear states. That is, the wave functions remain unchanged by adding a constant to the $T=0$ or 1 part of the interaction. This modification only affects the relative energies of the states with different isospin quantum numbers.

The matrix element for the $fpg$ model space, for which we took from Ref. \cite{hon09}, are defined in the particle-particle channel by assuming $^{56}$Ni as the core. The mass dependence of the interaction is assumed to be $(A/58)^{-0.3}=0.87$ for $A=92$ where $A$ is the mass number of the nucleus to be calculated. For $A=92$
we take the single-hole energies (relative to that of $0g_{9/2}$) of $\varepsilon(0g_{9/2})=0$ MeV, $\varepsilon(1p_{1/2})=0.144$ MeV, $\varepsilon(1p_{3/2})=1.443$ MeV and $\varepsilon(0f_{5/2})=4.407$ MeV.

\begin{figure}
\begin{center}
\includegraphics[scale=0.45]{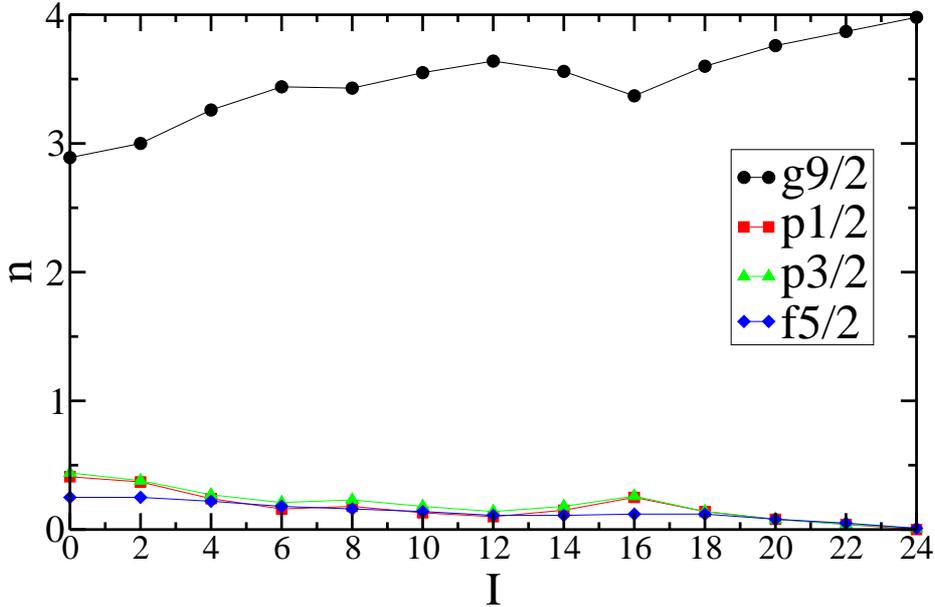}\\
\caption{Average number of proton (or neutron) holes in different orbitals for shell-model calculations on nucleus $^{92}$Pd in the $fpg$ shell \cite{hon09}.}
\label{pdnum}
\end{center}
\end{figure}

The calculated results for Ru and Cd isotopes are plotted in Fig. \ref{ru88}. Calculations are done with the shell-model code described in Ref. \cite{qi08}. Calculations for the spectra of Pd isotopes could be found in Refs. \cite{ced10,qi11}. Present shell model calculations are able to include a large number of shells with the help of modern computers. In Fig. \ref{pdnum} we calculated the average number of holes in each orbitals of the $fpg$ model space in which four orbitals are involved. It is thus seen that it is the $0g_{9/2}$ shell that dominates the occupancy. Also comparison with $pg$ and $0g_{9/2}$ calculations tend to suggest that many properties of  nuclei in this region can be explained by calculations restricted the the single $0g_{9/2}$ shell only. This does not mean that the other shells (the $0f1p$ orbitals and even other higher lying shells) has no contribution to the wave function. But normally one may safely expect the effect from these shells can be taken into account through the renormalization of the effective interaction and effective operators (c.f., Table \ref{table}). The contributions from other shells may be deemed as the background whose effect on the nuclear structure is hard to identify. This is because our knowledge of complex objects like atomic nuclei are obtained through systematic studies of neighboring states. As a result, only relevant degrees of freedom that distinguish one state from the others can be identified.
Simple models  are only for us to understand the physics and the residual degrees of freedom. The real situation can be much more complex. Even the shell-model single-particle orbitals are very complicated from a microscopic point of view and can be
very different from the real single-particle wave functions of the nucleus \cite{Talmi11,Talmi93}. Thus the shell occupancy plotted in Fig. \ref{pdnum} is a model dependent quantity. 
Also full $fp$ shell model calculation has been feasible for the nucleus $^{48}$Cr since almost 20 years ago. But it is realized that the $0f_{7/2}$ and $1p_{3/2}$ shells would be enough to explain the bulk properties of this deformed nucleus \cite{Juo00}. 

The regularly-spaced energy spectra in $^{92}${Pd} and $^{96}$Cd might naively
be interpreted as a collective vibrational motion. At present it is not clear how nuclear vibration can be described properly from a microscopic shell model perspective. 
On the other hand, it is suggested that the apparent collectivity in the spectrum of $^{92}$Pd may
have a different origin than that resulting from a vibration or rotational
motion usually attached to nuclear collective phenomena \cite{ced10,qi11}. 
Since there is no evidence for shell interferences in the spectra of Pd
isotopes, we will analyze other sources which control the structure of those
spectra, namely the different coupled pair modes. This can be done rather
easily in this case where the shell $0g_{9/2}$ is very dominant, by 
considering separately
the various components of the interaction. 
We found that when
the $T=0$ $np$ interaction is switched off, the corresponding
spectrum is of seniority type. This underlines the importance of the
$np$ interaction to induce collective
spectra. 

To probe the pair content in the many-fermion wave function $\Psi$ one may evaluate the two-particle transfer amplitude $\langle\Psi_N||(a_i^{\dag}a_j^{\dag})_{J^{\pi}}||\Psi_{N-2}\rangle$ or the average number of pairs $\langle\Psi_N||((a_i^{\dag}a_j^{\dag})_{J^{\pi}}\times(a_ia_j)_{J^{\pi}})_0||\Psi_{N}\rangle$ or project the shell-model wave functions onto a pair coupled basis with the help of the two-particle coefficients of fractional parentage. 
In Refs. \cite{ced10,qi11,qi11a} we applied the two-particle coefficients of fractional parentage technique. Two sets of the orthonormal bases are constructed starting from the monopole $J=0$ and $J=9$ pairs. In Refs. \cite{Qi10,qi10a,xu11} all possible combinations are considered within an non-orthogonal basis by applying the so-called multistep shell model.
After projecting the yrast wave functions into a product of isoscalar 
$J^\pi=9^+$ pairs we found the most striking feature of this case: \emph{all} low-lying yrast states
are built mainly from $J^\pi = 9^+$ spin-aligned $np$ pairs as 
$|((\nu\pi)_{J=9})^4_I\rangle$. 
In contrast to the isovector pairing excitations, in the
aligned isocalar pair state there is no pair breaking when the nucleus is
excited from one state to higher lying states in the spectrum.
Instead, angular momentum in
$^{92}$Pd is determined by a new form 
of coupling \cite{ced10}. The $((\nu\pi)_9)^N_I$ mode can be deemed 
as a \emph{collective} pair coupling scheme which can be written as a 
coherent superposition of all  isovector neutron and proton pairs occupying the $g_{9/2}$ shell in the 
form of $((\nu^2_J)^{N/2} \otimes (\pi_J^2)^{N/2})_I$. 
The dynamics of the system is determined by the active four
$(\nu\pi)_9$ hole pairs aligned differently according to the total angular
momentum of the system. 

In a restricted single $j$ shell, the calculations of the average pair number can be much simplified with the help of two-particle coefficients of fractional parentage as
\begin{equation}
C^I_J=\frac{n(n-1)}{2}\sum_{\alpha\beta}X_{\alpha\beta}M_{\alpha\beta}^I(J)X_{\beta},
\end{equation}
and
\begin{eqnarray}
\nonumber M^I_{\alpha\beta}(J) &=&\sum_{\alpha_{n-2}I_{n-2}}[j^{n-2}(\alpha_{n-2}I_{n-2})j^2(J)I|\}j^n\alpha I]\\
&&\times[j^{n-2}(a_{n-2}I_{n-2})j^2(J)I|\}j^n\beta I],
\end{eqnarray}
where $n$ is the number of particles (holes) of the system and $X$ are the expansion amplitudes of the wave function. The Greek letters denote the basis states and the summation runs over all possible states.

\begin{figure}
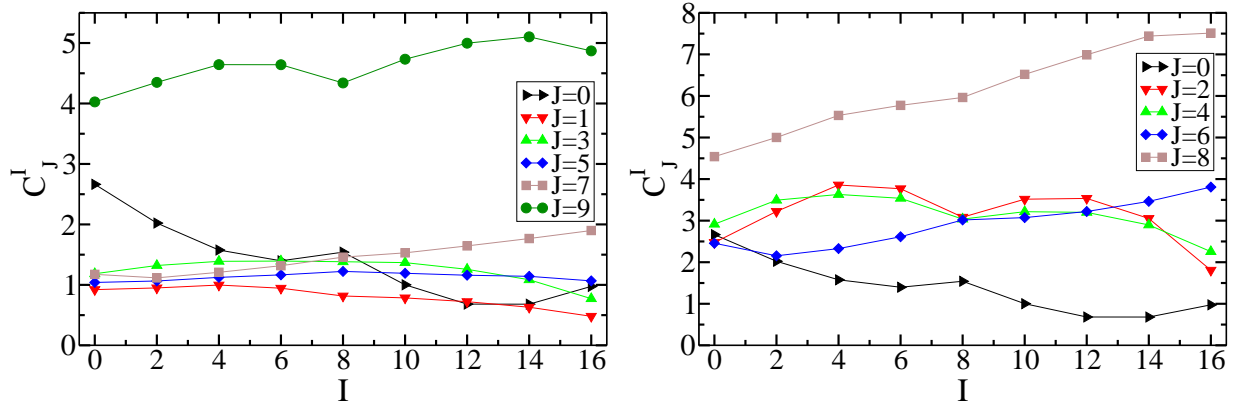

\begin{center}
\includegraphics[scale=0.33]{num.eps}~~ \includegraphics[scale=0.33]{isovec.eps}\\
\caption{Left: Average number of isoscalar $(0g_{9/2}^2)_J$ interacting pairs, $C^I_J$, as a function of total angular momentum $I$ for the wave functions of the yrast states of $^{92}$Pd; Right: Same as the left panel but for the numbers of isovector $(0g_{9/2}^2)_J$  pairs.}
\label{num}
\end{center}
\end{figure}

For systems with four pairs the total number of interacting pairs is
\begin{equation}
N=n(n-1)/2=28.
\end{equation} 
For a low-lying yrast state in $^{92}$Pd which has total isospin $T=0$, the total number of isoscalar pairs is 
\begin{equation}
[n/2(n/2+1)-T(T+1)]/2=10.
\end{equation} 
If isospin symmetry is assumed, the numbers of isovector neutron-neutron, proton-proton and $np$ pairs are the same and the total number of pairs is 
\begin{equation}
[3n/2(n/2-1)+T(T+1)]/2=18.
\end{equation} 

The results for the isoscalar pairs thus obtained in the $pg$ space are shown in the left panel of Fig. \ref{num}. One sees that \emph{all} low-lying yrast 
states are built mainly from $J^\pi = 9^+$ spin-aligned $np$ pairs.
In contrast to the isovector pairing excitations, in the
aligned isoscalar pair state there is no pair breaking when the nucleus is
excited from one state to higher lying states in the spectrum.
Instead, the angular momentum of excited yrast states in
$^{92}$Pd is generated by the rearrangement of the angular momentum vectors 
of the aligned $np$ pairs while the isoscalar pair character itself is 
preserved. For the ground state the four pairs are maximally aligned
in opposite directions as allowed by the Pauli principle. 
When exciting the nucleus from one yrast state to
another, the dynamics of the system is determined by the active four
$(\nu\pi)_9$ hole pairs aligned differently according to the total angular
momentum of the system. It should be mentioned that the microscopic origin of the equidistant pattern generated by the aligned $np$ pairs remain unclear at present. From a simple semiclassical point of view, one may argue that the corresponding energies
vary approximately
linearly with $I$ for small total angular momenta, inducing similar energy spacings 
between consecutive levels. 

To gain deeper physical insight into the structure of $^{92}$Pd we also calculated the numbers of isovector pairs which are plotted in the right panel of Fig.~\ref{num}. It is seen that already in the ground state we have much more pairs with $J>0$ than the normal $J=0$ pair. In particular, the dominating component is the $J=8$ pair which is maximally aligned in the isovector channel. This is due to the large overlap between states generated by the spin-aligned $np$ pairs and those generated by the isovector pairs. This phenomenon is not seen in systems with two $np$ pairs where the contributions from the isovector aligned pair is practically zero for low-lying yrast states \cite{Qi10}. It should be emphasized that the dominating component in the wave functions of low-lying yrast states in $^{92}$Pd is still the spin-aligned $np$ pair coupling. But it may be interesting to clarify the role played by the isovector aligned pair in $N=Z$ systems with more than two pairs. Work on this direction is underway.

\begin{figure}
\begin{center}
\includegraphics[scale=0.35]{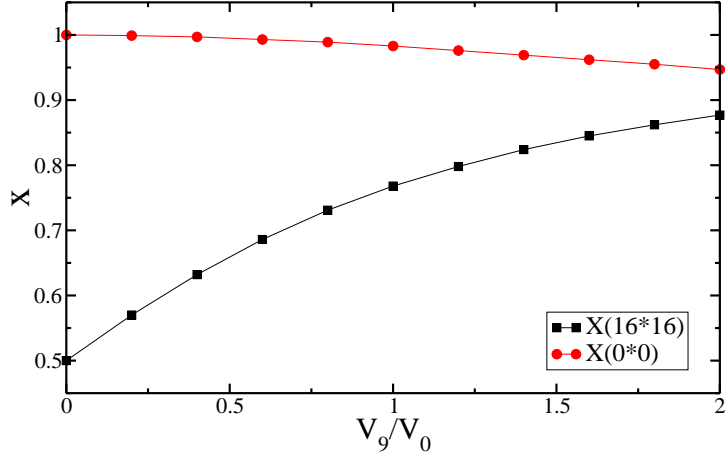}\\
\caption{ Wave function amplitudes for the components $|^{96}$Cd(gs)$\otimes $$^{96}$Cd(gs)$\rangle$ and $|^{96}$Cd($16^+$)$\otimes$$ ^{96}$Cd($16^+)\rangle$ in the ground state wave function of $^{92}$Pd  as a function of the ratio between the strengths of $V_9$ and $V_0$. The calculation is done in the $0g_{9/2}$ shell with all other matrix elements are set to zero.}
\label{pdwave}
\end{center}
\end{figure}

The four $J=9$ $np$ pairs in $^{92}$Pd can couple in various ways. With the help of two-particle cfp one may express the wave function in terms of $((((\nu\pi)_9\otimes(\nu\pi)_9)_{I'}\otimes(\nu\pi)_9)_{I''}\otimes(\nu\pi)_9)_{I}$.
It is thus found that, among the various aligned np pair configurations, the dominating components can be well represented by a single configuration $((((\nu\pi)_9\otimes(\nu\pi)_9)_{I'=16}\otimes(\nu\pi)_9)_{I''=9}\otimes(\nu\pi)_9)_{I}$. In the $0g_{9/2}$ shell, this configuration is calculated to occupy
around 66\% of the ground state wave function  of $^{92}$Pd, i.e., with amplitude $X(0^+_1)=0.81$ \cite{qi11}. 
Also it should be mentioned that the $J=9$ term is not the generator for the full aligned np coupling. It only generates the stretch configuration \cite{dan67} shown above.
The maximal $I=24$ state corresponds to a pure stretch configuration.
On the other hand, by rewriting the wave function of $^{92}$Pd as a product of two group with two $np$ pairs (i.e., $^{96}$Cd each) within an non-orthornormal basis it is thus found the leading component is coupling $|^{96}$Cd(gs)$\otimes $$^{96}$Cd(gs), as can also be seen from Fig. \ref{pdwave}.
This is consistent with the weak coupling calculations for light nuclei \cite{wong71}.

\begin{figure}
\begin{center}
\includegraphics[scale=0.35]{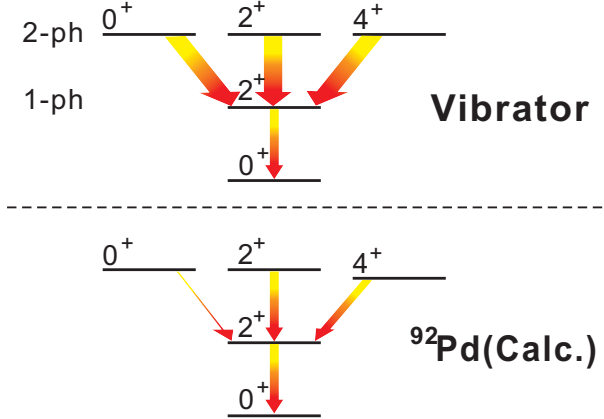}\\
\caption{Upper: The two-phonon triplet states predicted by the vibrational model. Lower: The calculated triplet of states in $^{92}$Pd with the relative $B$(E2) values (proportional to the width of the arrow) connecting them to the $2^+_1$ state. The calculation is done in the $fpg$ shell \cite{hon09}.}
\label{twoph}
\end{center}
\end{figure}

Usually the vibrational
spectrum is characterized by collective E2 transitions between states differing by one phonon,
where the corresponding transition strengths increase linearly as a function of the
spin of the initial state. One of the most important criteria to decide
whether a spectrum corresponds to the one of a vibrator is that there
should  be a nearly-degenerate two-phonon triplet with collective E2 
transitions to the one-phonon state $2^+_1$. In Fig. \ref{twoph} we plotted the low-lying level scheme of  $^{92}${Pd} predicted by $fpg$ shell model calculations with the interaction of Ref. \cite{hon09}. It is seen 
that there is indeed a triplet of states in 
the calculated spectrum which are nearly
degenerate and lie at twice the energy of the state $2^+_1$. 
The vibrational model predicts that the E2 transitions from the two-phonon
states to the state $2^+_1$ should all have the same strength and its value 
should be twice that of the $2^+_1\rightarrow 0^+_1$ strength.
This prediction differs very much from the calculated values for the aligned coupling scheme shown in the
Figure. In particular, the strength of the $0^+_2\rightarrow 2^+_1$
transition is one order of magnitude smaller than the predicted value for a vibrator. This may be due to the fact that the calculated second $0^+$ state contain large contributions from shells (in particular the $1p_{1/2}$ shell) other than the $0g_{9/2}$ shell. That is, this state does not correspond to the spin-aligned $np$ coupling scheme proposed in this work. Indeed, this $0_2^+$ state cannot be reproduced in calculations within the $0g_{9/2}$ shell.

The three yrast states in Fig. \ref{twoph} are mainly composed of the spin-aligned $np$ pairs \cite{qi11}. For example, the $2^+_1$ state is dominated by the configuration of  $^{96}{\rm Cd (gs)}\otimes^{96}{\rm Cd} (2^+_1)$ with $x=0.973$. The projection of this state on the stretch configuration $^{96}{\rm Cd} (16^+)\otimes^{96}{\rm Cd} (16^+)$ is $x=0.862$. It should be mentioned that the second $2^+$ state in $^{92}$Pd is also dominated by the spin-aligned $np$ pair coupling. In this state the largest two components are $^{96}{\rm Cd} (2^+_1)\otimes^{96}{\rm Cd} (2^+_1)$ with $x=0.904$ and $^{96}{\rm Cd} (14^+_1)\otimes^{96}{\rm Cd} (16^+)=0.766$.
This is in contrast with the coupling of $J=0$ pairs where only one state can be generated from the monopole pairs irrespective of the number of pairs involved.
An interesting question would be how many states can we construct from the aligned $np$ pairs. A detailed analysis will be done elsewhere.

\begin{figure}
\begin{center}
\begin{tikzpicture}[scale=0.45, y=1.0cm,x=1.0cm,every text node part/.style={font=\small} ]
\def\gsposy{0}
\def\firstbandx{0.6}
\def\levelwidth{1.6}
\def\bandwidth{3.0}
\def\nuclidespace{0.60}
\def\yscale{0.0010}
\def\nucposy{19.5}
\def\yscale{0.0140}
\def\x{\firstbandx}
\def\y{\gsposy+\yscale * \energy}
\foreach \energy / \sp in {0/$0^+$}{
\draw[thick] (\x,\y) -- (\x+\levelwidth,\y);
\node[above] at (\x+\levelwidth,\y) {\small  \energy};
\node[above] at (\x,\y) {\sp};
}
\node[below] at (\x+0.5 * \levelwidth,\gsposy+2.30) {\large Exp.};

\def\x{\firstbandx+1*\bandwidth}
\def\y{\gsposy+\yscale * \energy}
\foreach \energy / \sp in {0/$0^+$,614/$7^+$,976/$9^+$,925/$8^+$,843/$1^+$,841/$2^+$}{
\draw[thick] (\x,\y) -- (\x+\levelwidth,\y);
\ifthenelse{\energy = 841}{\node[above] at (\x+\levelwidth,\y-\yscale*70) {\small  \energy};}{\node[above] at (\x+\levelwidth,\y) {\small  \energy};}
\ifthenelse{\energy = 841}{\node[above] at (\x,\y-\yscale*70) {\sp};}{\node[above] at (\x,\y) {\sp};}

}
\node[below] at (\x+0.5 * \levelwidth,\gsposy+2.30) {\large fpg};

\def\x{\firstbandx+2*\bandwidth}
\def\y{\gsposy+\yscale * \energy}
\foreach \energy / \sp in {0/$0^+$,760/$7^+$,1109/$9^+$,870/$2^+$, 1064/$8^+$, 1142/$1^+$}{
\draw[thick] (\x,\y) -- (\x+\levelwidth,\y);
\ifthenelse{\energy = 1064}{\node[above] at (\x+\levelwidth,\y-\yscale*70) {\small  \energy};}{\node[above] at (\x+\levelwidth,\y) {\small  \energy};}
\ifthenelse{\energy = 1064}{\node[above] at (\x,\y-\yscale*70) {\sp};}{\node[above] at (\x,\y) {\sp};}

\ifthenelse{\energy=878}{\node[below] at (\x+0.5*\levelwidth,\y) {\large \emph{15}};}{}
\ifthenelse{\energy=1708}{\node[below] at (\x+0.5*\levelwidth,\y) {\large \emph{20}};}{}

}
\node[below] at (\x+0.5 * \levelwidth,\gsposy+2.0) {\large pg};

\def\x{\firstbandx+3*\bandwidth}
\def\y{\gsposy+\yscale * \energy}
\foreach \energy / \sp in {0/$0^+$,709/$7^+$,1149/$9^+$,836/$2^+$, 1037/$8^+$, 1082/$1^+$}{
\draw[thick] (\x,\y) -- (\x+\levelwidth,\y);
\node[above] at (\x+\levelwidth,\y) {\small \energy};
\node[above] at (\x,\y) {\sp};
\ifthenelse{\energy=878}{\node[below] at (\x+0.5*\levelwidth,\y) {\large \emph{15}};}{}
\ifthenelse{\energy=1708}{\node[below] at (\x+0.5*\levelwidth,\y) {\large \emph{20}};}{}

}
\node[below] at (\x+0.5 * \levelwidth,\gsposy+2.0) {\large g};

\end{tikzpicture}~~~~~~~\includegraphics[scale=0.32]{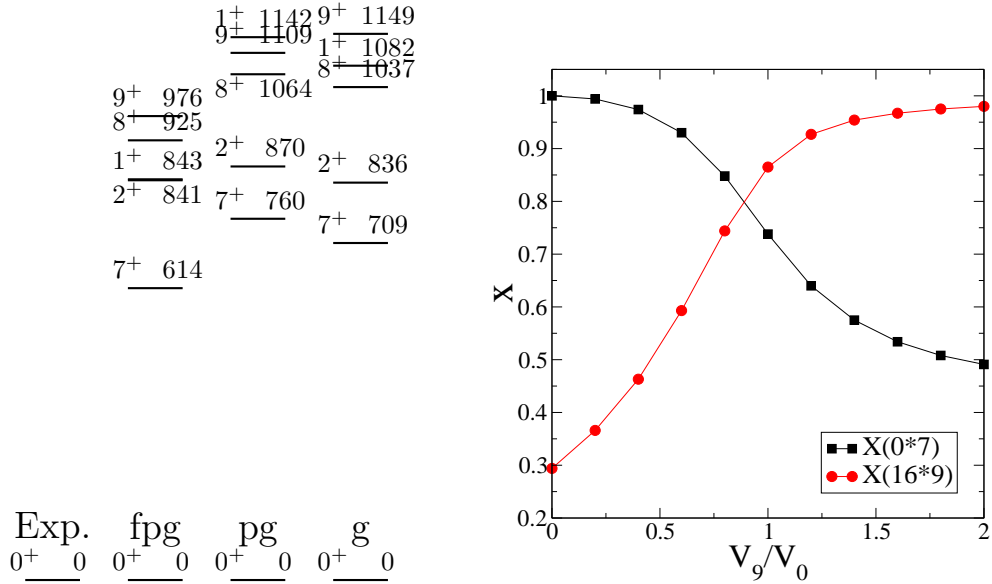}
\caption{Left: Experimental and calculated low-lying states in the odd-odd nucleus $^{94}$Ag.  Right: Wave function amplitudes for the components $|^{96}$Cd(gs)$\otimes (\nu\pi)_7\rangle$ and $|^{96}$Cd($16^+$)$\otimes (\nu\pi)_9\rangle$ in the $7^{+}_1$ state of $^{94}$Ag  as a function of the ratio between the strengths of $V_9$ and $V_0$. The calculation is done in the $0g_{9/2}$ shell with all other matrix elements are put as zero for simplicity.}
\label{ag94}
\end{center}
\end{figure}

The proposed spin-aligned $np$ coupling may also shed some new light on the structure of odd-odd $N=Z$ nuclei. Here we take the nucleus $^{94}$Ag as an example. It has three $np$ hole pairs in the $^{100}$Sn core. The calculated spectra of $^{94}$Ag in different model spaces are plotted in Fig. \ref{ag94}. The ground state is calculated to be $J^{\pi}=0^+$ and $T=1$ and is dominated by the coupling of two aligned $np$ pair and a normal isovector $J=0$ pair. The lowest $T=0$ state is calculated to be a $7^+$ state with excitation energy $E*\sim0.7$ MeV. This state is dominated by the aligned $np$ pair coupling of $(((\nu\pi)_9\otimes(\nu\pi)_9)_{I'=16}\otimes(\nu\pi)_9)_{I}$ with $|X|\approx 0.97$ in restricted calculations performed within the $0g_{9/2}$ shell. This component decreases to $|X|=0.83$ for the first $9^{+}$ state, which is otherwise dominated by the coupling $(((\nu\pi)_9\otimes(\nu\pi)_9)_{I'=0}\otimes(\nu\pi)_9)_{I}$. Again the wave functions of these states are mixtures of many components if we express them in the normal proton-proton and neutron-neutron coupling scheme.

In the right panel of Fig. \ref{ag94} we evaluated the wave function of the $7^{+}_1$ state in  $^{94}$Ag in the $0g_{9/2}$ shell with a schematic Hamiltonian containing two interaction terms with $J=0$ and $J=9$. It is thus found that the wave function shows a pure configuration of $|^{96}$Cd(gs)$\otimes (\nu\pi)_7\rangle$ for the calculation with a $J=0$ term only, while the stretch configuration $|^{96}$Cd($16^+$)$\otimes (\nu\pi)_9\rangle$  gradually takes over by enhancing the strength of the aligned pair interaction. Similar calculations were also done for the  $9^{+}_1$ state. For $V_9=0$ the wave function is characterized by the pure configuration $|^{96}$Cd(gs)$\otimes (\nu\pi)_9\rangle$ which is calculated to be the dominated component over a large range of $V_9/V_0$. The contribution from the stretch configuration $|^{96}$Cd($16^+$)$\otimes (\nu\pi)_9\rangle$ is also enhanced by enlarging the interaction $V_9$.

\begin{figure}[htdp]
\begin{center}
\includegraphics[scale=0.35]{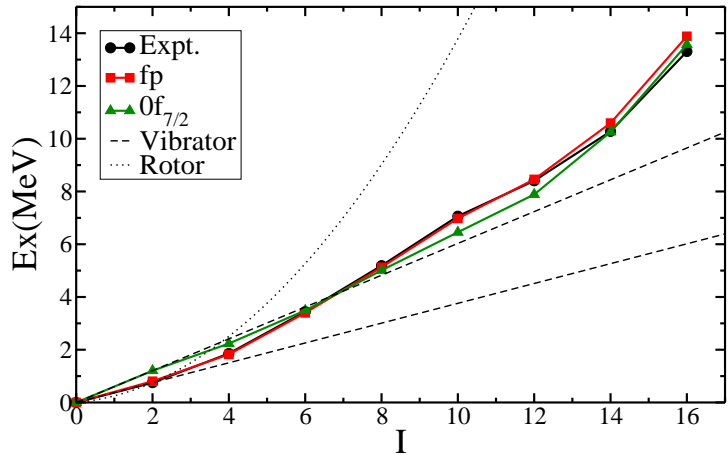}\\
\caption{Experimental \cite{Nudat} and calculated level schemes of the nucleus $^{48}$Cr. Calculations in the $fp$ and $0f_{7/2}$ shells are done with the interactions from Ref. \cite{kb3} and Ref. \cite{Esc}, respectively. The dashed and dotted lines correspond to the predictions of the the geometric collective model normalized to the $2^+_1$ states.}
\label{cr48}
\end{center}
\end{figure}

To gain deeper insight into the properties of the aligned $np$ pair structure and the regions of the 
nuclear chart where this exotic mode can be observed, we also performed calculations for the $N$, $Z=28$ mass region. It is 
noted that, when the calculations are restricted to the $0f_{7/2}$ shell only, both the $N=Z$ nuclei $^{44}$Ti 
and $^{48}$Cr exhibit an equally-spaced pattern similar to the one obtained with the shell $0g_{9/2}$, as can be seen in Fig. \ref{cr48}. 
However, the aligned $T=0$ np coupling is not manifested in the spectra of the $fp$ shell
since the proximity of the shell $1p_{3/2}$ and the strong
quadrupole interaction thus developed leads to the mixing of shell model
configurations that break the spin-aligned $np$ pair structure. As a result of this a
combination of $T=1$ pairing and quadrupole interaction forms a more favorable 
description and a deformed mean field is generated. To explore further the role played by the aligned $np$ pair in this shell, we calculated the yrast spectra of $^{48}$Cr with different strengths for the $J=7$ interaction. It is thus found that a transition from rotational-like spectrum to vibrational-like spectrum occurs by enhancing the strength of the $J=7$ interaction element. On the contrary, our calculations show that in the mass 90-100 region and the 
$0g_{9/2}$ shell the quadrupole 
interaction with the $1d_{5/2}$ level is not strong enough 
to scatter nucleons appreciably across the energy gap associated with the magic numbers
$N,Z=50$.  Work underway is to explore in a more quantitative way the influence of deformation on the aligned $np$ pair coupling.

Another ongoing work is to generalize the aligned $np$ pair coupling to systems with more than one shell. This generalization may not be straightforward within the shell model framework due to the fact that the corresponding aligned $np$ pairs can carry different angular momenta. One possibility to overcome this drawback is through introducing as building blocks four-body quartets which are composed of aligned $np$ pairs \cite{Danos69}.

\section*{Acknowledgments}
This work has been supported by the Swedish Research Council (VR) under grant 
Nos. 623-2009-7340 and 621-2010-4723. CQ thanks Prof. Honma for sending him the interaction matrix elements of the $fpg$ model space.

\end{document}